\documentclass[aps,prl,showpacs,twocolumn]{revtex4}
\usepackage{graphicx}
\begin{document}
\draft
\title{Equation of state and collective frequencies of a trapped Fermi gas along the BEC-unitarity crossover}
\author{G.E. Astrakharchik(a), R. Combescot(c), X. Leyronas(c) and S. Stringari(a,b)}
%\address{(a) Dipartimento di Fisica, Universit\`a di Trento and BEC-INFM, I-38050 Povo, Italy}
\affiliation{(a) Dipartimento di Fisica, Universit\`a di Trento and BEC-INFM, I-38050 Povo, Italy}
%\address{(b) Coll\`ege de France and Laboratoire Kastler-Brossel,
\affiliation{(b) Coll\`ege de France and Laboratoire Kastler-Brossel,
 Ecole Normale Sup\'erieure*, 
24 rue Lhomond, 75231 Paris Cedex 05, France}
%\address{(c) Laboratoire de Physique Statistique, Ecole Normale Sup\'erieure*,
\affiliation{(c) Laboratoire de Physique Statistique, Ecole Normale Sup\'erieure*,
24 rue Lhomond, 75231 Paris Cedex 05, France}
\date{Received \today}
%\maketitle

\begin{abstract}We show that the study of the collective oscillations in a harmonic trap provides a very sensitive test of the equation of state of a Fermi gas near a Feshbach resonance. Using a scaling approach, whose high accuracy is proven by comparison with exact hydrodynamic solutions, the frequencies of the lowest compressional modes are calculated at $T=0$ in terms of a dimensionless parameter characterizing the equation of state. The predictions for the collective frequencies, obtained from the equations of state of mean field BCS theory and of recent Monte-Carlo calculations, are discussed in detail.
\end{abstract}

\maketitle

%\begin{multicols}{2}

The availability of  Feshbach resonances and the consequent possibility of tuning the interatomic potential through the application of a magnetic field have recently stimulated much experimental and theoretical work on the BEC-BCS crossover in ultracold trapped Fermi gases \cite{levico}. Several relevant and already measured quantities, like the release energy, the density profiles and the frequency of the collective oscillations,  are related in a sensitive way to the equation of state of the homogeneous system  which varies significantly through the crossover. For small and negative values of the scattering length $a$ the equation of state approaches the limit of a non interacting Fermi gas while, for small and positive values, bound molecules can be formed and, at zero temperature, the system behaves like a dilute molecular Bose-Einstein condensed gas. In the intermediate regime  important theoretical issues  remain. For example the intermolecular scattering length on the BEC side exhibits a non trivial dependence on the free atom scattering length \cite{strina,gora}. Furthermore such interactions should give rise to beyond mean field corrections in the equation of state \cite{lhy}, not accounted for by  the Bogoliubov approximation. Similar corrections could also result from the composite nature of the molecules. Finally, near resonance, where the scattering length becomes larger than the average distance between particles, no simple many-body  approach is applicable to this strongly correlated system. Numerical calculations of the equation of state of a uniform interacting Fermi gas at T=0 along the crossover have been carried out in the past along the mean field BCS (MF-BCS) approach \cite{rander} and, more recently, through {\it ab initio} simulations based on Monte Carlo (MC) algorithms \cite{carl,astr}.

Since the  frequencies of the collective oscillations can be measured with high precision, it is of major interest to investigate their dependence on the equation of state along the crossover. From a careful and systematic  analysis of the frequencies one should be able to infer the actual equation of state. It has already been pointed out \cite{str2} that the collective frequencies  of a T=0 superfluid Fermi gas, trapped in a harmonic potential, approach well defined values in the important BEC and unitarity limit regimes, where the density dependence of the chemical potential can be inferred from general arguments. In the intermediate region various investigations, based on the use of the hydrodynamic theory of superfluids and suitable parametrizations of the equation of state, have appeared recently \cite{hu,heisel,bulbert,kim,mani,comm,modmod}.  In the mean time first experimental results  \cite{thomas,grim} on the frequencies of the lowest axial and radial compression modes on ultra cold gases of $^{6}$Li across the Feshbach resonance have also become available.

The purpose of this paper is to establish the accuracy needed in the measured frequencies in order to determine experimentally the equation of state. With this goal in mind we will focus on two specific theories for the equation of state which exhibit a rather different density dependence of the chemical potential through the crossover. These are the MC calculations and the MF-BCS approximation discussed above. These equations of state are then employed within a hydrodynamic approach in order to calculate the relevant collective frequencies. Hydrodynamic theory is justified at T=0 by the superfluid nature of the system and  has been already  successfully employed in trapped Bose-Einstein condensed gases. We will restrict here the discussion to the case of positive scattering
lengths, where the experimental conditions for achieving the superfluid hydrodynamic regime are less severe, in contrast with the BCS regime where the smallness of the gap gives rise to stringent constraints \cite{comm}. The determination of the collective frequencies, for a given equation of state, is by no means a trivial task because, in general, the solutions must be found numerically in the non symmetric configurations relevant for experiments. Furthermore since the frequencies vary only moderately with the scattering length, the analysis should be carried out with high precision. In the following we will discuss only  the lowest relevant  modes, namely the monopole mode for spherical geometry and the lowest axial compression mode for cigar configurations.

Let us first start discussing the two equations of state and, in particular, the dependence on density $n$ of the chemical potential $\mu (n)$. Instead of the density, we use the Fermi wavevector $k_F$ defined by $ k^{3}_{F}=3 \pi ^{2} n$. We restrict ourselves to the case of a wide Feshbach 
resonance where the scattering length $a$ is the only relevant length associated with the two-body interaction, as it is indeed the case \cite{fesh} for the dominant resonance in $^{6}$Li and also for 
$^{40}$K. This leads to introduce the dimensionless coupling parameter $1/k_F a$, which varies from zero to infinity when one goes from  unitarity to the bosonic limit of molecules. Similarly we express the chemical potential in units of the Fermi energy $E_F= \hbar^{2}k^{2}_{F}/2m$. The equations of state are displayed in Fig. 1.
%\vspace{-7mm}
\begin{figure}[htbp]
\begin{center}
%\scalebox{0.5}{\rotatebox{270}{\includegraphics[width=10cm]{figMCMFBCSmu2}}}
\scalebox{0.7}{\includegraphics[width=10cm]{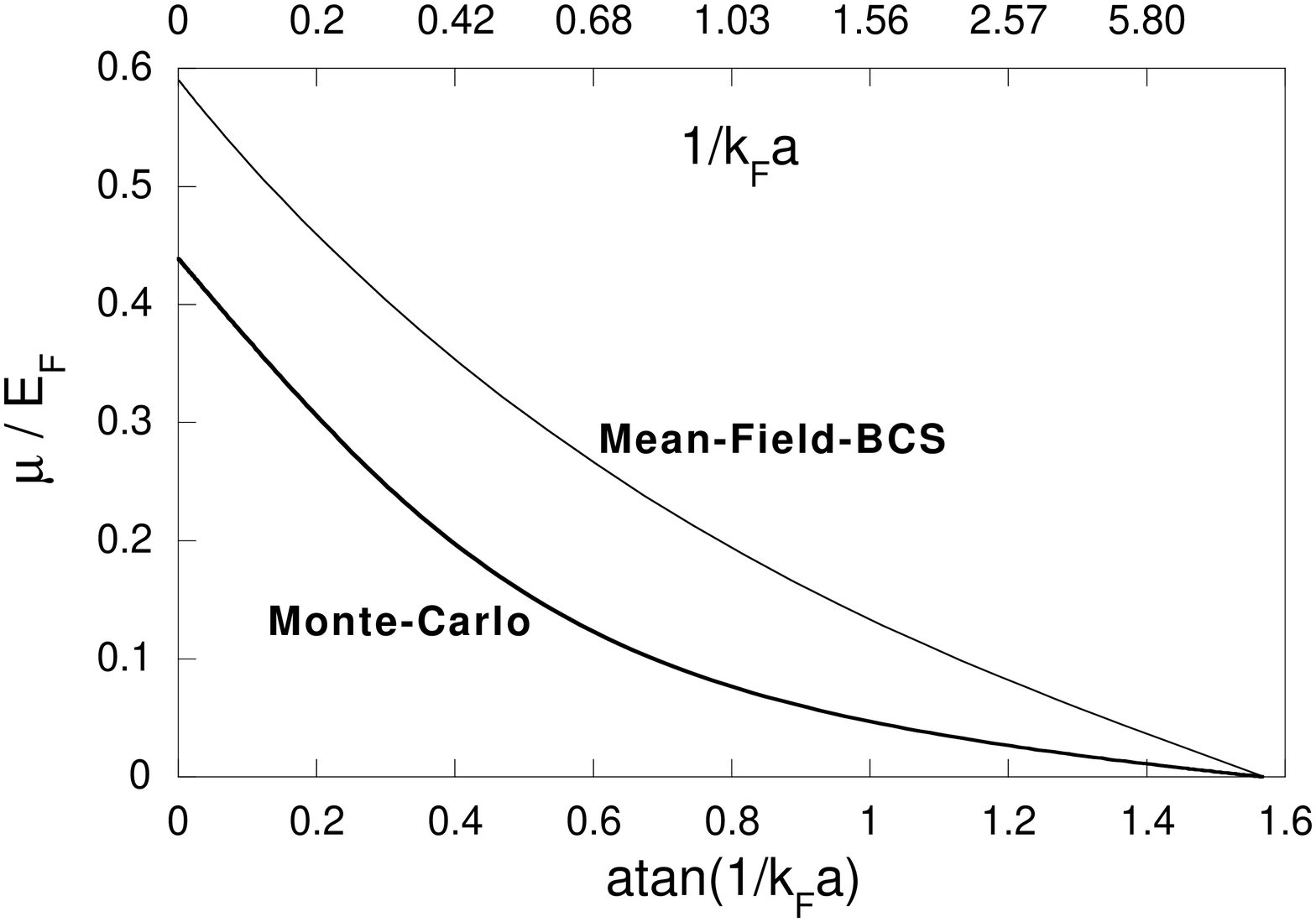}}
\caption{Chemical potential $\mu (n)$(shifted by half the molecular binding energy $\hbar^{2}/2ma^{2}$) in units of the Fermi energy $E_F$ as a function of the coupling parameter $1/k_Fa$. The Monte-Carlo data are extracted from Ref. [7].}
\label{figmu}
\end{center}
\end{figure}
\vspace{-5mm}

At unitarity the result $\mu/E_F = 0.44$ extracted from MC calculations \cite{carl,astr} is in fair agreement with experiments \cite{bourdel,grimm}, while the MF-BCS theory gives a larger  result $\mu/E_F = 0.59$. The slopes near unitarity are very similar, but the approach to the BEC limit is quite different reflecting the fact that MC calculations are consistent with the recent prediction  $ a_M=0.6 \,a$ for the molecule-molecule scattering length \cite{gora}, while the MF-BCS \cite{rander} yields the significantly larger value $a_M=2 a$. 

In order to evaluate the frequencies of the collective modes, different routes have been followed in the literature.  One class of methods assumes that the equation of state of the gas can be closely approximated by a polytropic expression $\mu (n) \simeq n^{\gamma}$ and makes use of the analytical results known \cite{cozz} for this case. Note that the polytropic approximation is exact in the BEC regime where $\mu=gn$, as well as at unitarity where dimensionality considerations yield $\mu\propto n^{2/3}$. Other approaches are based on exact numerical solutions of the hydrodynamic equations \cite{fuchs}. Here we will consider another method based on the use of  scaling transformations. Scaling has been already successfully employed to describe the dynamical behaviour of Bose-Einstein condensates \cite{castdum,shl}. We will see that the results of the scaling approach agrees remarkably well with exact solutions of hydrodynamic equations in all cases of practical interest we have considered. The advantage of the proposed method is its full applicability to anisotropic traps where exact numerical solutions are not always available. On the other hand it can be also applied to regimes where the equation of state is not of the polytropic form.

In our scaling approach  we use the following  Ansatz for the space dependence of the density distribution
\begin{eqnarray}
n(x_1,x_2,x_3,t) = \gamma(t) \,n_0(a_1(t)\,x_1,a_2(t)\,x_2,a_3(t)\,x_3)
\label{eqscal}
\end{eqnarray}
where $n_0({\bf r})$ is the equilibrium density and 
$\gamma(t)= \prod_{i} a_i(t)$. The factor $\gamma(t)$ ensures that the total particle number $\int d{\bf r} \,n({\bf r},t) $ is conserved.
Correspondingly, we take $v_{s,i}({\bf r})=-(\dot{a}_i/a_{i}) x_i$ for the superfluid velocity field which ensures that the equation of continuity  $ \dot{n}({\bf r}) + {\bf \nabla} \left[ n({\bf r}) \,{\bf v}_{s}({\bf r}) \right] = 0$ is automatically satisfied. We will then insert  the above Ansatz for the density and for the velocity field into the hydrodynamic Lagrangian 
\begin{eqnarray}
{\mathcal{L}}=  \int d{\bf r} \left[ \frac{1}{2}\, m\,n({\bf r}) \,{\bf v}_{s}^{2}({\bf r}) - e(n({\bf r})) - n({\bf r}) V_h({\bf r}) \right]
\label{lagr}
\end{eqnarray}
of our zero temperature superfluid. Here $m$ is the atomic mass and the chemical potential $\mu (n)$ is linked to the energy per unit volume $e(n)$ by $\mu (n)=\partial e(n)/\partial n$. 
The density dependent energy term $e(n)$ in the Lagrangian has been here obtained in the Local Density Approximation, consistently with the hydrodynamic description.

In the following we will specialize  to the case of small oscillations, suited to the collective modes in the linear regime. Writing $a_i=1+\epsilon _i$ where $\epsilon _i \ll 1$, in the equations of motion obtained from the Lagrangian (\ref{lagr}), we have to carry out the calculation up to first order in $\epsilon _i$.
After some algebra we find the following simple eigenvalue equations for the mode frequencies $\omega $:
\begin{eqnarray}
\Gamma \,\omega ^{2}_{i}  \sum_{j} \epsilon _j + 2\, \omega ^{2}_{i} \,\epsilon _i =  \omega ^{2} \,\epsilon _i
\label{eqmod}
\end{eqnarray}
where we have naturally used $\ddot{\epsilon }_i=-\omega ^{2}\epsilon _i$ and we have defined
the relevant dimensionless parameter 
\begin{eqnarray}
\Gamma = \frac{3}{2}\, \frac{\langle n \,\frac{\partial \mu }{\partial n} \rangle}{\langle V_{ho} \rangle}-1
\label{eqxi}
\end{eqnarray}
where  $V_{ho}({\bf r}) =  \sum_{i} (1/2) m \omega ^{2}_{i} x^{2}_{i}$ is the trapping harmonic potential with  $x_1=x$, $x_2=y$ and $x_3=z$. 
The brackets are for the equilibrium average over the gas cloud, for example $\langle V_{ho} \rangle=  \int d{\bf r} \,n_0({\bf r})\,V_{ho}(\bf {r})$  and the density profile is evaluated in the 
Thomas-Fermi approximation $\mu(n)+V_{ho}({\bf r})=\mu_0$, consistently with the hydrodynamic description.  Notice that the parameter $\Gamma$ does not depend on the anisotropy of the trap. Using the virial theorem it is easy to prove that, for a polytropic equation of state $\mu(n)\propto n^\gamma$, the parameter $\Gamma$ coincides with the exponent 
$\gamma$ so that at unitarity one has $\Gamma=2/3$, while in the BEC regime $\Gamma=1$.  In general the coefficient $\Gamma$ is a function of the dimensionless  combination $k_F(0)a$ where $k_F(0)$ is fixed by the central value of the Thomas-Fermi density distribution according to $ k^{3}_{F}(0)=3 \pi ^{2} n(0)$. The actual dependence of $\Gamma$ on $k_F(0)a$ is determined only by the equation of state $\mu(n)$. Finally the above derivation can be also reformulated as starting from an exact variational principle for the hydrodynamic frequencies.

In the isotropic case $\omega _i = \Omega $, in addition to the quadrupole mode $\omega ^{2}=2 \Omega ^{2}$, Eq.(\ref{eqmod}) gives the result $\omega ^{2}=(3\Gamma+2) \,\Omega ^{2}$ for the monopole mode  and, in particular, 
 the well-known result $\omega =\sqrt{5} \,\Omega$ in the BEC limit $\Gamma=\gamma=1$. In the axisymmetric geometry $\omega _1=\omega _2=\omega _{\perp}, \omega _1=\omega _z$, we find, in addition to the quadrupole mode $\omega ^{2}=2 \omega_{\perp} ^{2}$, the two $m=0$ compressional solutions: 
\begin{eqnarray}
\omega ^{2} = \frac{1}{2} \; [ \;2 (\Gamma +1) \,\omega ^{2}_{\perp} + (\Gamma+2)\, \omega ^{2}_{z} \hspace{25mm} \nonumber \\
\pm  \sqrt{ \left[ 2(\Gamma +1) \,\omega ^{2}_{\perp} - (\Gamma+2) \,\omega ^{2}_{z} \right] ^{2} + 8 \Gamma^{2}\,\omega ^{2}_{\perp} \,\omega ^{2}_{z} } \;\; ]
\label{eqmod1}
\end{eqnarray}
For the relevant case of cigar geometry $\omega_z \ll \omega_{\perp}$ one finds
 $ \omega ^{2} = \omega ^{2}_{z} (3-1/(\Gamma +1) )$ for the lowest axial compression  mode, giving  $ \omega/ \omega_{z}= \sqrt{5/2}$ in the BEC limit. For the radial compressional mode one finds instead $ \omega ^{2} = 2(\Gamma +1) \,\omega ^{2}_{\perp}$.

We will now show that the scaling result is extremely accurate by comparing its predictions
with essentially exact solutions of the hydrodynamic equations. Our method \cite{fuchs} makes use of known exact analytical results for a large class of model equation of states $\mu_{anal} (n)$. Then the actual $\mu (n)$ is very closely approximated by one member of this class. Moreover the effect of the small difference between $\mu (n)$ and $\mu_{anal} (n)$ is taken into account using a perturbative approach (this is called the "corrected" model). The absolute precision of this method has been checked \cite{fuchs} to be of order $10^{-3}$ at least. The relevant ingredients needed to calculate the mode frequencies are the reduced gas density in the trap $ \bar{n}({\bf r}) = n({\bf r})/n_0$ and the corresponding chemical potential $ \bar{\mu }({\bf r}) = \mu ({\bf r})/\mu_0$, both normalized to their central values $n_0$ and $\mu_0$. A class of  soluble models is described by the law $ \bar{n} = \bar{\mu }^{p} \exp{ [\sum_{k=0}^{K} P_{k}\bar{\mu } ^{k}]}$, where $p$ and $P_{k}$ are parameters. Here we consider only the cases $K=0$ (which is the polytropic model $\bar{n} =\bar{\mu }^p$) and $K=1$, because they give already by far a good enough precision. The model $K=1$ is called "quasi-polynomial". In this last case the solution is not fully analytical, but it involves only finding the proper root of a low order polynomial, which is a quite trivial numerical task \cite{fuchs}.

We have carried out the calculations of the collective frequencies in two different physical situations. The first one is the isotropic geometry which is particularly convenient for deriving exact solutions of the hydrodynamic equations. The second one is the elongated cigar geometry employed in many experimental set-up. Note  that the results $\nu^{2}= 4$ for the unitary case and $\nu^{2}= 5$ for the BEC limit are much more  separated in the spherical case than in the cigar geometry where one finds instead, for the lowest axial mode, $\nu^{2}= 2.4$ and $\nu^{2}= 2.5$ respectively.
\vspace{-5mm}
\begin{figure}[htbp]
\begin{center}
%\scalebox{0.5}{\rotatebox{270}{\includegraphics[width=10cm]{figMCMFBCS+diff2}}}
\scalebox{0.7}{\includegraphics[width=10cm]{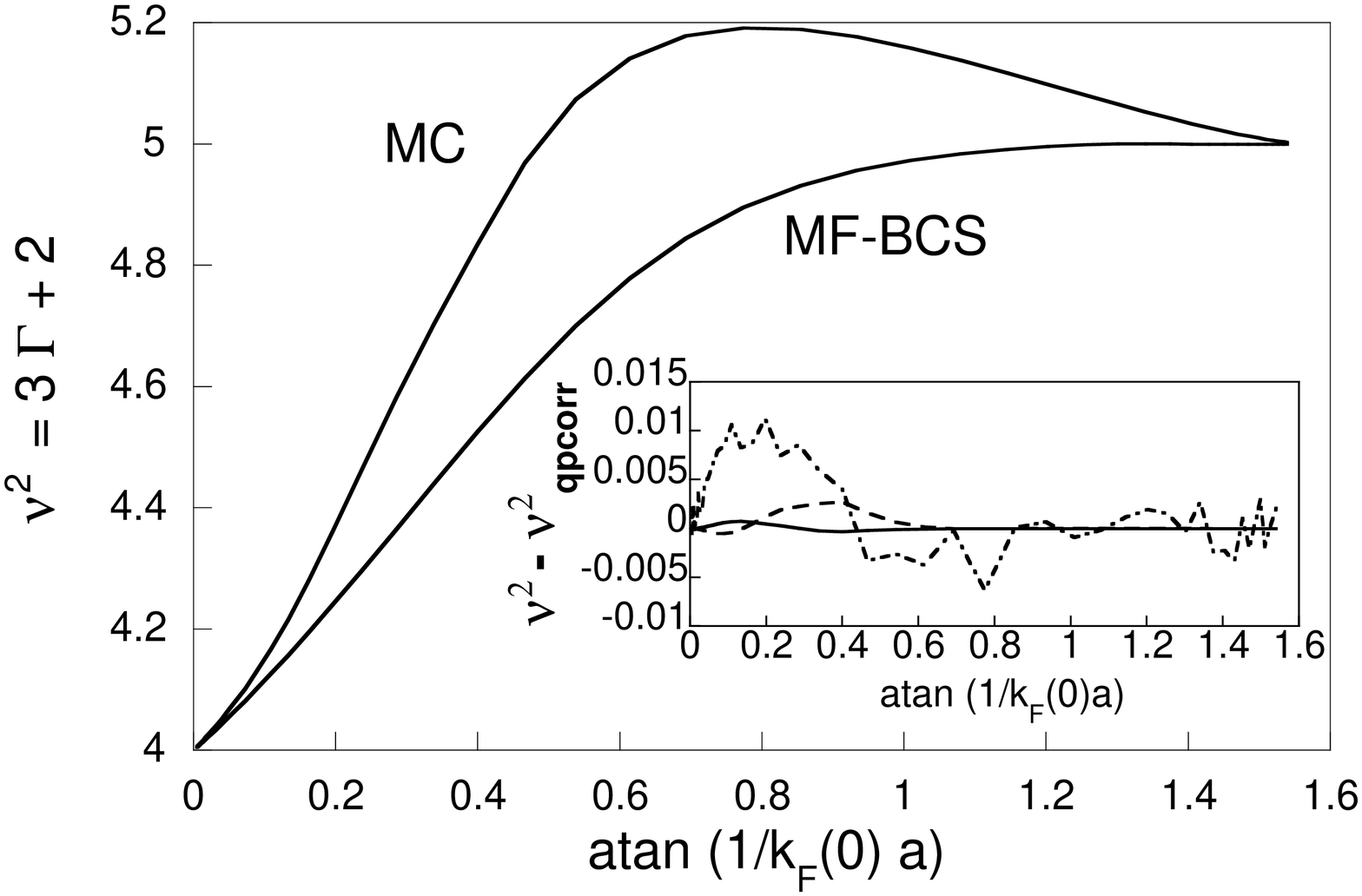}}
\caption{Reduced square $\nu^{2}=\omega ^{2}/\Omega^{2}$ of the lowest monopole frequency in isotropic trap for the Monte-Carlo (MC) and the mean-field BCS (MF-BCS) equation of states as a function of $1/k_F(0) a$, where $k_F(0)$ is the Fermi wavevector at the trap center. Full lines corresponds to the results of all our methods, which can not be distinguished at this scale. Insert : (uncorrected) quasi-polynomial (dotted-dashed line), corrected polytropic (full line) and scaling (dashed), all compared to the corrected quasi-polynomial model.}
\label{figmcbcs}
\end{center}
\end{figure}
\vspace{-8mm}

We display  our results in Fig. 2 for the spherical geometry. We have considered  the two equations of state provided by MC calculations and by the MF-BCS approximation. The accuracy of the scaling approach  is in both cases remarkable since at the scale of this figure it coincides with the exact result. The excellent quality of the scaling approach can be understood because it is exact not only for the polytropic case, but also if one includes first order corrections to the polytropic equation of state \cite{note}, as can be proven from the variational formulation mentioned above.
In order to display the precision of our results, we have  plotted the  results in the insert of Fig.2 on a much larger scale, taking the MC equation of state as an example. More precisely we have taken as a reference the frequency $\nu_{qpcorr}^{2}$ obtained by the corrected quasi-polynomial method, which we believe is the most accurate. Indeed we see first that the corrected polytropic model and the corrected quasi-polynomial are in remarkable agreement, since the difference between them is at most $8. 10^{-4}$. Hence the convergence of our method toward the exact result is almost already achieved at the level of the simpler corrected polytropic model. This clearly proves that the much more precise corrected quasi-polynomial model is certainly extremely accurate, since it is without correction already within at most $10^{-2}$ of the exact result. The next striking  and important result is the quite remarkable accuracy of the scaling method which is essentially never beyond $3. 10^{-3}$ of the exact result for the MC equation of state (the case of the MF-BCS is even better). This gives a full validation to the scaling approach.
\vspace{-3mm}
\begin{figure}[htbp]
\begin{center}
%\rotatebox{270}{
\scalebox{0.7}{\includegraphics[width=12cm]{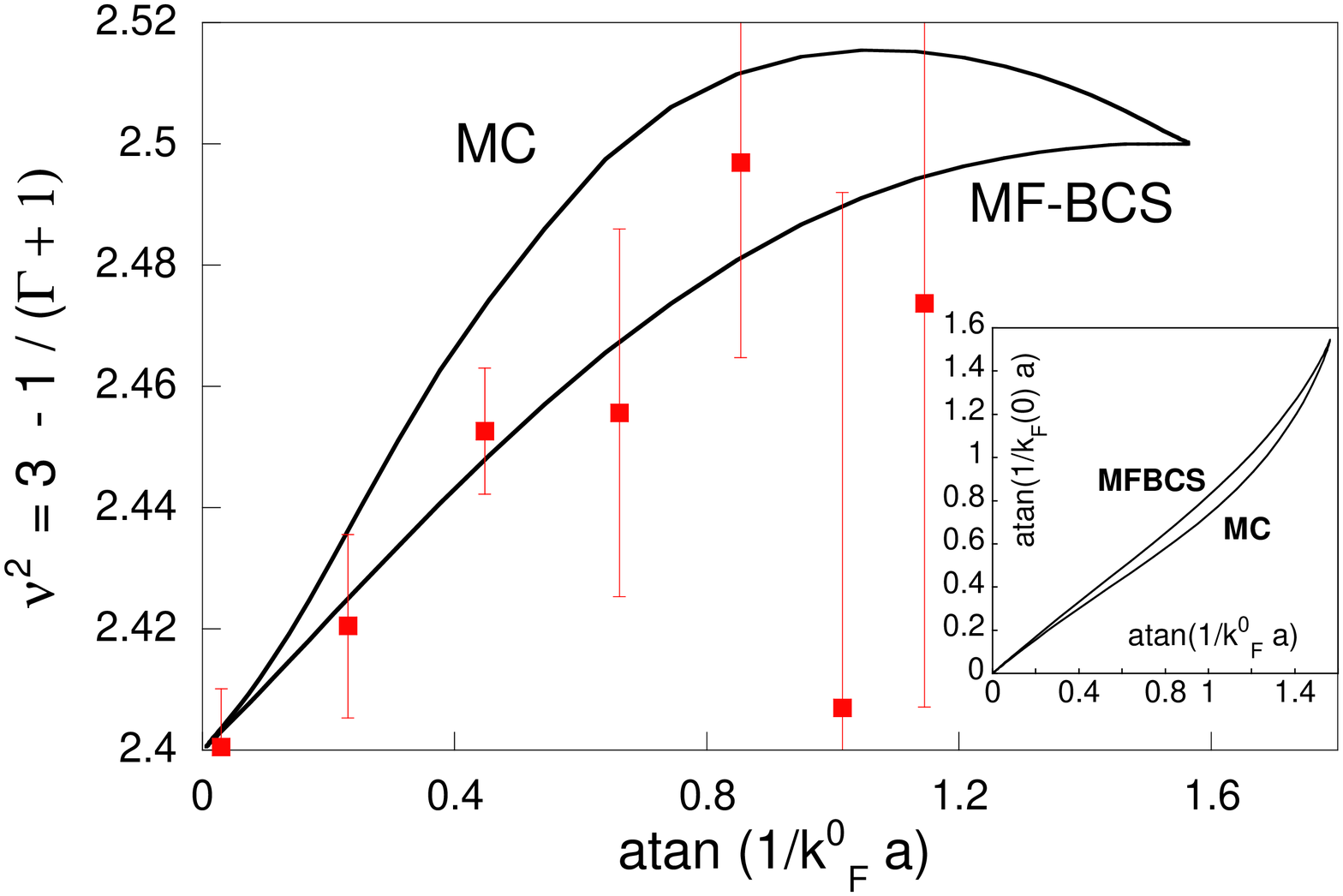}}
%}
\caption{Reduced square $\nu^{2}=\omega ^{2}/\omega^{2}_z$ of the axial mode frequency in very elongated trap for the Monte-Carlo (MC) and the mean-field BCS (MF-BCS) equation of states, as a function of $k_{F}^{0}a$, where $k_{F}^{0}=1.7 N^{1/6}/a_{ho}$ is the Fermi wavevector of the ideal Fermi gas in a harmonic trap and $a_{ho}=(\hbar /m{\bar \omega} )^{1/2}$ where ${\bar \omega}= (\omega _x \omega _y\omega _z)^{1/3}$.  Experimental results are from Ref. [10]. The insert gives the relation between $k_{F}^{0}$ and $k_{F}(0)$, used in Fig. 2.}
\label{figcigar}
\end{center}
\end{figure}
\vspace{-8mm}

An essential feature emerging from the results of Fig. \ref{figmcbcs} is that the separation between the MC and the MF-BCS results is definitely much larger than any theoretical uncertainty within the $T=0$ hydrodynamic picture. It should be consequently possible to distinguish them experimentally. In particular the observation of an enhancement of $\nu^{2}$ above 2.5, for $k_{F}^{0} a \sim 1$, would provide evidence for beyond mean-field corrections \cite{lhy,ps} which are absent in MF-BCS.

Finally we turn to the cigar geometry. We have carried out an analysis of the lowest axial compressional mode by comparing  the results obtained from the scaling method with our exact results, in the case studied in Ref. \cite{modmod}. Again the agreement is perfect. In Fig. \ref{figcigar} we report the predictions of the scaling approach applied to the MC and the MF-BCS equations of state. In the same figure we also show the experimental results from \cite{grim} for the axial mode \cite{note2}. It is  quite intriguing that these data  look in better agreement with the mean-field BCS equation of state rather than with  the Monte-Carlo one. This might be due to finite temperature effects. Clearly further experimental work would be important to settle this point, which is essential for our understanding of the BEC-BCS crossover, and in particular to decide which  equation of state  better describes the experimental reality.

We are grateful to M. Bartenstein for sending us his data and C. Salomon for discussions. We acknowledge support from MIUR.

\noindent
*Laboratoire associ\'e au Centre National
de la Recherche Scientifique et aux Universit\'es Paris 6 et Paris 7.
%\begin{references}

%\end{multicols}

\begin{thebibliography}{99}
\bibitem{note2} The analysis can be extended to the radial compression mode (higher frequency solution of Eq.(\ref{eqmod1})). However, within the available experimental conditions,  the applicability of the hydrodynamic description to this 'high frequency' mode  is not fully guaranteed near unitarity due to the smallness of the gap.
\bibitem{note} On the other hand far away from this physical situation, scaling is less accurate. For example for the $\alpha -p$ model \cite{fuchs} , where $\bar{\mu } = 1 - (1-\bar{n}^{1/p})^{2/\alpha }$, scaling gives $\omega ^{2}/\Omega ^{2}=9/(4p)+5/4$ for $\alpha =1$, compared to the exact result $2/p+1$.
\end{thebibliography}
\end{document}